\documentclass[aps,pra,twocolumn,superscriptaddress,nofootinbib,showpacs]{revtex4}
\usepackage{amssymb}
\usepackage{graphicx}
\usepackage{epsf,epsfig}
\usepackage{bm}
\usepackage{bbm}
\usepackage{amsfonts}
\usepackage{amsmath}
\usepackage{graphics}
\usepackage[normalem]{ulem}

\newcommand{\be}{\begin{eqnarray}}
\newcommand{\ee}{\end{eqnarray}}
\newcommand{\ba}{\begin{array}}
\newcommand{\ea}{\end{array}}
\newcommand{\no}{\nonumber}

\newcommand{\eps}{\varepsilon}

\newcommand{\bfr}{{\bf r}}

\newcommand{\bfn}{{\bf n}}

\newcommand{\bfp}{{\bf p}}

\newcommand{\bfv}{{\bf v}}

\newcommand{\bfj}{{\bf j}}

\newcommand{\bflambda}{{\bm \lambda}}

\begin{document}

\title{Kinetics of the disordered Bose gas with collisions}
\author{G. Schwiete}
\email{schwiete@zedat.fu-berlin.de} \affiliation{Dahlem Center for Complex Quantum Systems and Institut f\"ur Theoretische
Physik, Freie Universit\"at Berlin, 14195 Berlin, Germany
}
\author{A. M. Finkel'stein}
\affiliation{Department of Physics and Astronomy, Texas A\&M University, College Station, TX 77843-4242, USA}
\affiliation{Department of
Condensed Matter Physics, The Weizmann Institute of Science, 76100
Rehovot, Israel}
\date{\today}

\begin{abstract}
We discuss the kinetics of the disordered interacting Bose gas using the Boltzmann transport equation. The theory may serve as a unifying framework for studying questions of dynamics of the expanding Bose gas at different stages of the expansion. We show that the transport theory allows us to straightforwardly reproduce and generalize a number of results previously obtained from microscopic models in different formalisms. Based on estimates for the inter-particle scattering rates, we discuss the relevance of interaction effects for the localization problem in the interacting disordered Bose gas. We argue that, if the number of particles is large enough, the size of the expanding cloud may exceed the localization length. We describe the spreading of the wave packet in this regime as collision-induced diffusion and compare the obtained rate of expansion to known results on subdiffusive spreading in non-linear disordered lattices.
\end{abstract}

\pacs{71.10.Ay, 71.10.Pm, 75.40.Cx} \maketitle

\section{Introduction} 

Studies of localization effects in disordered Bose gases are at the forefront of research on cold atomic gases \cite{Clement05,Fort05,Lye05,Schulte05,Billy08,Roati08,Hulet09,Dries10,Robert10,Jendr12}. In a typical setup, the gas is released from a trap and expands under the influence of a disorder potential. The evolution of the expanding gas cloud is often divided into two stages \cite{Shapiro07}. At the first stage, the density is comparatively high, the interparticle interaction dominates, and disorder-related effects are neglected. At the following stage, the density of the cloud drops so that the average kinetic energy (per particle) becomes much larger than the interaction energy. For this second stage, the phenomenon of Anderson localization for non-interacting particles is studied. As we discuss in this paper, this picture is not complete due to the effects induced by interparticle collisions.

In Refs.~\onlinecite{Schwiete10, Schwiete10a} we raised the question of how the presence of interactions influences the expansion of a matter wave-packet in a disorder potential. During the course of the expansion, the energy stored in an interaction-induced Hartree potential gradually converts into kinetic energy, and the combined effect of disorder and interactions leads to interesting dynamics. This study as well as the related work of Ref.~{\onlinecite{Cherroret11}} were based on the Gross-Pitaevskii equation (GPE), thereby assuming a large occupation of participating modes that allows to describe the Bose gas as a non-linear medium. Initially, only the collision-less limit was considered, while the effect of collisions has been studied only recently \cite{Schwiete13}.

In this work, building upon insights gained from the microscopic analysis of the GPE, the appropriate kinetic equation for the disordered interacting Bose gas is obtained from a semi-phenomenological Boltzmann-equation approach. The underlying physics is essentially classical: Throughout the paper we assume that the interaction energy is sufficiently small compared to the kinetic energy so that one may neglect off-diagonal components of the Bogoliubov Hamiltonian.

The classical theory of nonlinear diffusion based on the Boltzmann equation can serve as a starting point for a qualitative analysis of the effect of interparticle interactions on localization. There are three main effects. First, the time-dependent interaction-induced potential and the inter-particle collisions can both lead to dephasing. In particular, the inter-particle collision rate is a lower bound for the dephasing rate when considering effects of localization. Second, for an out-of-equilibrium distribution, collisions lead to a particle-flux down in energies. Since particles with a low energy are much more likely to localize, this process may serve as a seed for localization. Third, the wave packet may continue to spread as a result of interparticle collisions despite the fact that single particle states are localized. The findings can be important for experiments on disordered Bose gases as soon as interactions are not completely negligible on the stage of localization.

This paper is organized as follows. In Sec.~\ref{sec:Boltzmann} we derive an effective nonlinear diffusion equation starting from the semi-phenomenological Boltzmann equation. We consider systems for which disorder scattering is the most effective scattering mechanism, while inter-particle collisions are sufficiently infrequent.
In the resulting equation, the interaction-induced collision integral simplifies considerably since in the diffusive limit the distribution functions may be taken as isotropic.
This allows us to provide analytic expressions for the kernel of the collision integral in the two- and three-dimensional cases.
We point out that for a two-dimensional gas the mean squared radius is a linear function of time, whenever the impurity scattering can be modeled by a white noise random potential. This result can be used as a reference point for numerical simulations.
Next, we make contact with our recent work based on the GPE in the limit of large occupation numbers \cite{Schwiete13}. In Sec.~\ref{sec:loc} we discuss the spreading of the wave packet resulting from interparticle collisions.
We predict that if the number of particles is large enough, the size of the expanding cloud may exceed the localization length of the noninteracting theory (e.g., evaluated at a typical energy of the cloud particles). We conclude in Sec.~\ref{sec:Conclusion}. In the Appendix, we demonstrate the effectiveness of the kinetic equation approach by reproducing with this method the diagrammatic results obtained in a work on thermalization of the disordered Bose gas for low occupation numbers, Ref.~\onlinecite{Geiger12}.

\section{Boltzmann equation approach} 
\label{sec:Boltzmann}

We consider the Boltzmann equation in the form
\be
&&\partial_t f_{\bfp}(\bfr,t)+\bfv\nabla  f_{\bfp}(\bfr,t)-\nabla U(\bfr)\nabla_{\bfp} f_{\bfp}(\bfr,t)\no\\
&&=C^d_{22}[f]+C_{imp}[f].\label{eq:Boltzmann}
\ee
Here, $\bfv=\bfp/m$ is the velocity, $f_\bfp$ the distribution function that is normalized as $n=\int (d\bfp) f_{\bfp}$, where $n$ is the density of particles and we use the abbreviation $(d\bfp)=d\bfp/(2\pi)^d$ in $d$ dimensions. The potential $U$ in general comprises two parts, a smooth classical external potential, and a self-consistent Hartree-like contribution, $\vartheta(\bfr,t)=2 g_d n(\bfr,t)$ \cite{Schwiete10,Cherroret11}. The coupling constant in three dimensions equals $g_3=4\pi a_s/m$, where $a_s$ is the scattering length; in two dimensions one should set $g_2=g_3/a_z$, where $a_z$ is the length which characterizes the transverse confinement. The interpretation of the above equation is a statistical one: We consider quantities that are averaged over many realizations of disorder. This is why the density $n$, for example, is a smooth quantity. The equation should be supplemented with initial conditions, alternatively one may add a source term to the right hand side of the equation.

Depending on the experimental setup, specifics of the system such as the finite correlation length of the disorder in a speckle potential need to be taken into account \cite{Kuhn07,Piraud12}. For example, the smooth external potential $U$ may incorporate the Fourier components of the speckle potential that are much smaller than $\lambda^{-1}$ (here $\lambda$ is the typical wavelength of the Bose-gas particles), while the shorter components are comprised into the impurity collision integral $C_{imp}$ on the right-hand side of the above equation. For the sake of simplicity, we will consider a simple relaxation type approximation, for which
\be
C_{imp}[f]=-\frac{1}{\tau_{p}}\left(f_\bfp(\bfr)-\left\langle f_{\bfp}(\bfr)\right\rangle_{\bfn}\right)
\ee
and $\left\langle\dots\right\rangle_{\bfn}$ denotes an averaging over the solid angle of the momentum. $\tau_p$ is the transport scattering time. The collision integral $C_{22}$ for the interacting Bose gas with a short-range pseudo-potential reads \cite{Griffin09},
\be
C^d_{22}[f]&=&\frac{4\pi g_d^2}{(2\pi)^{2d}}\int d\bfp_2d\bfp_3d\bfp_4 \;\delta(\bfp+\bfp_2-\bfp_3-\bfp_4)\no\\
&&\qquad\qquad\times \delta(\eps_p+\eps_{p_2}-\eps_{p_3}-\eps_{p_4})\mathcal{F}[f],\label{eq:C22}
\ee
where
\be
\mathcal{F}[f]=(1+f_{\bfp}+f_{\bfp_2})f_{\bfp_3} f_{\bfp_4}-f_{\bfp}f_{\bfp_2}(1+f_{\bfp_3}+f_{\bfp_4})\label{eq:occupfactors}
\ee
is a combination of distribution functions characteristic for the in- and out-scattering processes. For the sake of brevity, we suppressed the space arguments for the distribution functions.
It is worth mentioning that in the limit of large occupation numbers, i.e., when the GPE as a classical field equation is valid, $\mathcal{F}$ can be replaced by 
\be
\mathcal{F}_{GP}[f]=(f_{\bfp}+f_{\bfp_2})f_{\bfp_3}f_{\bfp_4}-f_{\bfp}f_{\bfp_2} (f_{\bfp_3}+f_{\bfp_4}). \label{eq:FGP}
\ee
In the context of wave turbulence, the collision integral with $\mathcal{F}_{GP}$ is derived based on a statistical averaging, resulting in a random phase approximation \cite{Zakharov84a,Zakharov92}. The collision integral in this form was also re-derived recently in a microscopic study of the expansion of a matter wave-packet based on the GPE \cite{Schwiete13}. Another limit of interest is the limit of low occupancy (LO), $f_{\bfp}\ll 1$, $f_{\bfp_i}\ll 1$. In this case, $\mathcal{F}$ may be replaced by 
\be
\mathcal{F}_{LO}[f]=f_{\bfp_3}f_{\bfp_4}-f_{\bfp}f_{\bfp_2}.\label{eq:FLO}
\ee

\subsection{Diffusive limit} 

We are interested in a stage of expansion of the Bose gas for which the average kinetic energy, $\overline{\eps}$, is already the largest energy scale. In particular, we will assume that $\overline{\eps}$ is much larger than the potential energy $\sim gn(\bfr)$. As we will see below, under this condition the scattering rate due to the inter-particle interaction, $1/\tau_{col}$, is much smaller than $\overline{\eps}$. (In other words, we can speak about states with a well defined energy.) Besides, we will assume that $\overline{\eps}$ is much larger than the disorder induced scattering rate $1/\tau_{p}$, where $\bfp$ is within the range of typical momenta, which determine $\overline{\eps}$. This inequality not only allows to develop a comprehensive theory of diffusion but, among other things, insures that in dimension $d=2$ the gas cloud expands for a long time before localization effects become essential. We address effects of collisions on localization in Sec.~\ref{sec:loc} below.

We focus on the regime for which disorder is the main scattering mechanism, $1/\tau_{p}\gg 1/\tau_{col}$. It can be expected that under these conditions the momentum is randomized on time scales of the order of the transport scattering time induced by the impurities and, therefore, the zeroth angular harmonic $f_0=\left\langle f_{\bfp}\right\rangle_{\bfn}$ is the dominant one. This motivates the standard ansatz used in the transport theory, 
\be
f_{\bfp}(\bfr,t)=f_0(p,\bfr,t)+d\;\bfn {\bf f}_1(p,\bfr,t),
\ee
where $d$ is the dimensionality, $p=|\bfp|$, $\bfn=\bfp/p$ and $|{\bf f}_1|\ll f_0$. Inserting this ansatz, a set of two equations can be derived by projecting the kinetic equation onto the zeroth and first angular harmonic,
\be
\partial_t f_0+v\nabla {\bf f}_1-\nabla^k U\left(\partial_p {\bf f}_1^k+\frac{d-1}{p}{\bf f}_1^k\right)=\left\langle C_{22}[f]\right\rangle_{\bfn},\\
\partial_t{\bf f}^i_1+\frac{v}{d} \nabla^i f_0-\nabla^i U\;\frac{1}{d}\partial_p f_0=-\frac{{\bf f}_1^i}{\tau_p}+\left\langle \bfn^i C_{22}[f]\right\rangle_{\bfn}.
\ee
Neglecting the small $\partial_t{\bf f}^i_1$ and $\left\langle \bfn^i C_{22}[f]\right\rangle_{\bfn}$ (recall that $\tau_p$ is the shortest time scale), one may resolve for 
\be
{\bf f}_1^i=-\frac{\tau_p v}{d}\left(\nabla^i -\nabla ^iU \partial_{\eps_p}\right)f_0.
\ee 
Here, and from now on, we assume that the energy spectrum is quadratic, $\eps_p=p^2/2m$. Then, entering the first equation, one gets
\be
&&\partial_t f_0-\left(\nabla -\nabla U\partial_{\eps_p}\right){D}_{\eps_p}\left(\nabla -\nabla U\partial_{\eps_p}+\Gamma_{\eps_p}\nabla U\right) f_0\no\\
&&=\mathcal{I}_d[f_0], \label{eq:diffusion}
\ee
where $D_{\eps}=2\eps\tau/(dm)$ is the diffusion coefficient, $\Gamma_{\eps}=\frac{2-d}{2\eps}=-\partial_{\eps}\ln \nu(\eps)$, $\nu(\eps)$ is the density of states and $\mathcal{I}_d[f_0]=C^d_{22}[f_0]$. Indeed, after entering with the ansatz for $f_{\bfp}$ into the averaged collision integral $\left\langle  C_{22}[f]\right\rangle_{\bfn}$, the largest contribution comes from those terms in which all factors of $f_{\bfp}$ are replaced by $f_0$. In the next section, we will provide explicit expressions for the collision integral $\mathcal{I}_d$. Equation~\eqref{eq:diffusion} has a rather broad range of applicability, it can be used for both stationary and for dynamical problems, for small as well as for large occupation numbers and also in different dimensions. Below in Sec.~\ref{subsec:relation}, we make contact with our previous studies of the diffusive nonlinear kinetic equation originating from the Gross-Pitaevskii equation.

In Appendix \ref{app:therm} we demonstrate how a recent result on the thermalization of a \emph{rarified} gas during a \emph{stationary} diffusion  process, Ref.~\onlinecite{Geiger12}, can be reproduced in a compact way as a limiting case of Eq.~\eqref{eq:diffusion}. For a review of recent progress on the general problem of non-equilibrium dynamics and thermalization see Ref.~\onlinecite{Polkovnikov11}.

\subsection{Collision integral} 

Since $f_0$ depends only on the modulus of $\bfp$, but not on $\bfn$, the angular integrations over $\bfn_2$, $\bfn_3$ and $\bfn_4$ in the expression for $\left\langle  C_{22}[f_0]\right\rangle_{\bfn}$ may be performed using a method devised in Ref.~\onlinecite{Hohenegger09}. The expression for $\mathcal{I}_d[f_0]$ can be written in the form 
\be
\mathcal{I}_d[f_0]&=&C_d \int d\eps_2d\eps_3 d\eps_4 \;\delta(\eps_p+\eps_2-\eps_3-\eps_4)\no\\
&&\times \chi_d(p,p_2,p_3,p_4)\mathcal{F}[f_0].\label{eq:I_d}
\ee

This result is obtained in the following way. Starting point is Eq.~\eqref{eq:C22}. After replacing $f\rightarrow f_0$ the angular averages in $\bfn_2$, $\bfn_3$ and $\bfn_4$ can be performed. To this end let us define
\be
R_d =\int d\bfn_2 d\bfn_3 d\bfn_4 \;\delta(\bfp+\bfp_2-\bfp_3-\bfp_4).\label{eq:S1}
\ee
After rewriting 
\be
\delta(\bfp+\bfp_2-\bfp_3-\bfp_4)=\int (d\bflambda)\mbox{e}^{i{\bflambda}(\bfp+\bfp_2-\bfp_3-\bfp_4)},\label{eq:S2}
\ee
one finds for the angular average
\be
R_d={\Omega_d}\int_0^\infty \frac{d\lambda \lambda^{d-1}}{(2\pi)^d}\eta_d(\lambda p)\eta_d(\lambda p_2)\eta_d(\lambda p_3)\eta_d(\lambda p_4).\label{eq:S3}
\ee
Here, $\Omega_d$ is the solid angle in $d$ dimensions, and $\eta_d(\lambda p)=\left\langle \mbox{e}^{\pm i {\bflambda} \bfp}\right\rangle_{\bfn}$. Further, 
\be
\eta_3(x)=\frac{\sin(x)}{x},\quad \eta_2(x)=J_0(x),\label{eq:S4}
\ee
where $J_0$ is a Bessel function of the first kind \cite{Abramowitz72}. In three dimensions, the remaining integral in $\lambda$ is elementary, while for two dimensions it can be found in Ref.~\cite{Nicholson20}. This brings us directly to Eq.~\eqref{eq:I_d}. What remains is to specify the Kernels $\chi_d$ and normalization constants $C_d$. The additional constraint imposed by the energy conservation in the collision integral leads to a simple form for the kernel $\chi_3$:
\be
\chi_3(p,p_2,p_3,p_4)&=&\sqrt{{\min(\eps_p,\eps_2,\eps_3,\eps_4)}/{\eps_p}} ,\label{eq:chi3}
\ee
where $p_{i}\equiv p_{\varepsilon _{i}}=\sqrt{2m\eps_{i}}$. The normalization constant for $d=3$ is $C_3=g_3^2m^3/2\pi^3$. In two dimensions, the kernel $\chi_2$ can be found as follows ($\Theta$ is the Heaviside function):
\be
\chi_2(p,p_2,p_3,p_4)&=&\Theta_{p_3+p_4-|p-p_2|}\Theta_{p+p_2-|p_3-p_4|}\label{eq:chi2}\\
&\times&\left\{\ba{cc}K\left({\Lambda}/{\Delta}\right)/\Delta,\quad \Delta\ge\Lambda\\K\left({\Delta}/{\Lambda}\right)/\Lambda,\quad \Delta<\Lambda\ea\right.,\no
\ee
where $\Lambda^2=pp_2p_3p_4$ and
\be
16 \Delta^2=\Pi_{i=1}^4 \left.(p_1+p_2+p_3+p_4-2p_i)\right|_{p_1=p}.
\ee
The function $K$ is the complete elliptic integral of the first kind. The kinematic constraints ensure that each of the factors on the right-hand side in the definition of $\Delta^2$ is non-negative. The normalization constant is $C_2=g_2^2 m^3/\pi^3$.

\subsection{Mean squared radius}

Although the diffusive kinetic equation (\ref{eq:diffusion}) constitutes a considerable simplification compared to the Boltzmann equation (\ref{eq:Boltzmann}), it is still difficult to solve.
It turns out, however, that for the special case of a two-dimensional system with a constant transport scattering time and for $U=\vartheta$ an analytical result for the time dependence of the mean squared radius of the evolving gas can be found
\be
\partial_t \left\langle \bfr^2_t\right\rangle=2dD_{\eps_{tot}},\label{eq:dtrsq}
\ee
where $\eps_{tot}=E_{tot}/N$ is the total energy per particle. This relation has already been noted in Refs.~\cite{Schwiete10,Schwiete13} for theories starting from the GPE. Here, we generalize this result to the range of applicability of the Boltzmann equation.

The reasoning is based on the conservation laws for particle number and energy. It is crucial for the arguments to consider the self-induced potential only, $U=\vartheta$.
The continuity equation for the density is easily obtained from the Boltzmann equation upon integration in $\bfp$: 
\be
\partial_tn+\nabla \bfj=0,\label{eq:S5}
\ee
where $n=\left\langle f_{\bfp}\right\rangle_\bfp$ and $\bfj=\left\langle f_{\bfp} \bfv\right\rangle_{\bfp}$ and $\left\langle \dots\right\rangle_{\bfp}$ means $\int (d\bfp)(\dots)$.
To obtain the continuity equation for the energy density, 
\be
\partial_t \rho_E+\nabla {\bf j}_E=0,\label{eq:S6}
\ee
one multiplies the Boltzmann equation by $\eps_\bfp$ and integrates in $\bfp$. It gives
\be
\rho_E=\left\langle \eps_{\bfp} f_{\bfp}\right\rangle_{\bfp}+g_dn^2,\quad {\bf j}_E=\left\langle \bfv \eps_{\bfp} f_{\bfp}\right\rangle_{\bfp}+\vartheta {\bf j}.\label{eq:S7}
\ee
We immediately conclude that the total energy $E_{tot}=\int d\bfr \rho_{E}$ is conserved. The mean radius squared is defined as 
\be
\left\langle \bfr_t^2\right\rangle=\frac{1}{N}\int d\bfr \bfr^2 n(\bfr).
\ee
From the continuity equation for the density it follows that 
\be
\partial_t\left\langle \bfr^2_t\right\rangle=\frac{2}{N}\int d\bfr \bfr {\bf j}(\bfr,t).
\ee 
Now consider the mean radius squared in the diffusive limit, for which ${\bf j}=\left\langle v {\bf f}_1\right\rangle_{\bfp}$. Two conditions are needed for the argument:  $(i)$ $\Gamma_{\eps}=0$, i.e., the density of states is constant, and $(ii)$ $D_{\eps}=k\eps$, where $k$ is a constant. Under these conditions ${\bf j}=-k\nabla \rho_E$. Plugging this into the relation for $\left\langle \bfr^2_t\right\rangle$ one immediately finds relation \eqref{eq:dtrsq} with $\eps_{tot}=E_{tot}/N$. Both conditions $(i)$ and $(ii)$ are fulfilled for the two-dimensional case with energy-independent $\tau$.

\subsection{Diffusion equation for the Gross-Pitaevskii equation} 
\label{subsec:relation}
Let us establish a connection with the diffusive kinetic equation that was previously derived microscopically from the GPE, first in the collisionless limit \cite{Schwiete10,Schwiete10a,Cherroret11}, and then in the presence of collisions \cite{Schwiete13,RemarkCollisions}. Upon setting $n'_\eps=f_0(p_\eps\equiv \sqrt{2m\eps})$ and $2\pi\nu(\eps) n'(\eps)=n_\eps$ in Eq.~(\ref{eq:diffusion}), one immediately arrives to the equation
\be
&&\partial_t n_\eps-\left(\nabla -\nabla U\partial_{\eps}-\Gamma_{\eps}\nabla U\right){D}_{\eps}\left(\nabla -\nabla U\partial_{\eps}\right) n_\eps\no\\
&&=2\pi\nu(\eps)\mathcal{I}_d[n'],\label{eq:GP}
\ee
where the collision integral contains $\mathcal{F}$ as defined in Eq.~(\ref{eq:occupfactors}), which is valid for arbitrary occupation numbers. Equation \eqref{eq:GP} is still equivalent to Eq.~\eqref{eq:diffusion}, we only passed from the momentum to the energy basis. The kinetic equation of Ref.~\onlinecite{Schwiete13} is recovered in the limit of large occupation numbers, $n'(\eps_i)\gg 1$, for which one may substitute $\mathcal{F}\rightarrow \mathcal{F}_{GP}$. Alternatively, the collision integral obtained by this replacement, $\mathcal{I}^{GP}$, can be derived microscopically starting directly from the GPE, see Ref.~\onlinecite{Schwiete13}. Accordingly, when one formally studies the solutions of the disordered GPE without reference to the Bose gas problem, Eq.~\eqref{eq:GP} with the collision integral $\mathcal{I}^{GP}$ remains applicable even in the limit of small occupation numbers.

%
%

\subsection{Collision rate}

The collision integral contains two terms describing the in and out collision channels. To estimate the scattering rate $1/\tau_{coll}$ of a state with a typical energy $\eps$, let us focus on the out term, which corresponds to the last term in the expression for $\mathcal{F}[f]$, see Eq.~(\ref{eq:occupfactors}). We will now be interested in the GP-limit of large occupations. Naturally, the out term is proportional to $n_{\eps}$;
we will write it as $n_{\eps}/\tau_{coll}$. Let us denote the typical kinetic energy per particle at point $\bfr$ as $\overline{\eps}(\bfr)$. For a conservative estimate of the scattering rate, let us consider an energy $\eps\sim\overline{\eps}(\bfr)$. Since one has to integrate a product of two distribution functions in the last term of Eq.~\eqref{eq:FGP}, this ultimately yields a factor $n^2({\bfr},t)$ in the scattering rate:
\be
\frac{1}{\tau_{coll}}\sim g_d^2 \frac{n^2({\bfr},t)}{\overline{\eps}(\bfr)}.\label{eq:colrate}
\ee
We consider the case when $g{_d}n(\bfr,t)\ll\overline{\eps}(\bfr)$ that leads to $1/\tau_{coll}\ll\overline{\eps}(\bfr)$.

\section{Localization and collision-induced diffusion} 
\label{sec:loc}

The problem of pulse propagation with large kinetic energy differs in many aspects from the flow of a superfluid system in the presence of disorder \cite{Huang92,Giorgini94,Lopatin02,Gaul11}. As is well known, for a Bose gas confined in a given volume, the redistribution induced by collisions leads to a seeding of a state with macroscopically large occupation at the bottom of the energy band \cite{Kagan92, Kagan95}. In contrast, in the case of a disordered medium, particles that get into states with energies $\eps_p\lesssim 1/\tau_p$ as a result of collisions are prone to localization. Let us denote the energy for which $\eps_p=1/\tau_p$ as $1/\tau ^{\ast}$. One may try to estimate the rate of generation of the density of localized particles by integrating the in-scattering term in the collision integral over the interval of energies $\eps\lesssim 1/\tau ^{\ast}$, i.e., $dn_{loc}/dt\approx dn/dt|_{\varepsilon \lesssim 1/\tau ^{\ast}}$. The analysis shows, however, that in three dimensions for the expanding cloud the mechanism of seeding of localized particles as a result of interparticle collisions is not very effective.

The situation in restricted dimensions is more involved. The difference from three dimensions is that in $d\leqslant 2$ \emph{all} states are localized. In two dimensions the scale of the localization length of noninteracting particles, $l_{loc}(\eps)$, increases exponentially with energy $\eps$ when it is larger than $1/\tau^*$. In a quasi one-dimensional geometry, as it is well known, \cite{Dorokhov83, Efetov83} $l_{loc}\sim N_{ch}l$, where $l$ is the disorder scattering length at $\eps \sim \overline{\varepsilon }$, and $N_{ch}\sim A/\lambda a_{\perp }\gg1$ is the number of channels in a wire or stripe with a cross-section area $A$; $a_{\perp }=\min(a_z,\lambda)$. (The strictly one-dimensional case will not be discussed; $a_{\perp}=a_z$ corresponds to a stripe.) Let us now consider a Bose-gas cloud of $N$ particles in $d\leqslant 2$ which can be characterized by a radius $R(t)$ at a time $t$. Roughly speaking, during the course of expansion, the states with the energy $\eps$ are available for localization at the time $t_{\eps}$ when $R(t_{\eps})$ becomes $\gtrsim l_{loc}(\eps)$. This, however, does not necessarily mean that such particles are indeed localized at this time. For localization, which implies a well defined state, it is necessary that the energy uncertainty should be smaller than the level spacing $\Delta$. Hence, for a particle with a typical kinetic energy $\overline{\eps}$, to start localizing at time $t_{\overline{\eps}}$, the condition
\be
\frac{1}{\tau_{col}(t_{\overline{\eps}})}<\Delta_{loc}=\frac{1}{\nu_{d} l^d_{loc}(\overline{\varepsilon })}
\ee
should be fulfilled; here $\nu_d$ is the effective density of states. This condition is similar to the comparison of the Thouless
energy (i.e., the uncertainty of the state) with the level spacing introduced by Thouless as a criterion for the metal-insulator transition. A necessary condition for localization of the main population of particles into single-particle states can be formulated as follows:
\be
\left\langle R^{2}(t)\right\rangle \gtrsim \max \left[{l^2_{loc}(\overline{\varepsilon }),R^{2}_{\Delta_{loc}}(d=1,2)}\right],\label{eq:radius}
\ee
where 
\be
R^{2}_{\Delta_{loc}}(2d)&=&\frac{a_s}{a_z}\lambda l_{loc}(\overline{\varepsilon })N,\\
R^{2}_{\Delta_{loc}}(1d)&=&\left(\frac{a_s}{a_{\perp}}\right)^2\lambda lN^2.
\ee
Thus, if the number of particles is large enough, e.g., in one dimension when $N/N_{ch}\gg (a_{\perp}/a_s)(l/\lambda)^{1/2}$, the expanding cloud will pass $l_{loc}(\overline{\varepsilon })$ without being stopped. The frequent collisions will make the interference, which is responsible for localization, ineffective \cite{CommentMotional}.

After reaching $R^{2}_{\Delta_{loc}}$, the expansion of the cloud does not halt. There are two mechanisms acting in parallel at $t \gtrsim t_{\overline{\eps}}$. One is the expansion of the cloud due to the collision-induced diffusion, with the diffusion coefficient \cite{Gogolin83}
\be
D_{col} \sim l^2_{loc}(\overline{\varepsilon })/ \tau_{col} \sim l^2_{loc}(\overline{\varepsilon }) g^2_{d} n^2 (\bfr,t)/\overline{\varepsilon }.\label{eq:Dcol}
\ee
Note that there is a broad interval where, despite the localization, the estimate \eqref{eq:colrate} for the collision rate derived from the Boltzmann collision integral is still valid. It ceases to work only when this rate becomes comparable with the \emph{many-body} level spacing \cite{Gornyi05,Basko06,Minkov07,Aleiner10}. The other mechanism is releasing potential energy stored in the cloud which generates a force $-\nabla \vartheta (\bfr,t)$ per particle.
As a result the cloud will expand according to a power-law rate $\left\langle R^{2}(t)\right\rangle \sim t^{\alpha}$, with some $\alpha < 1$. Let us discuss $\alpha$ for various mechanisms and different dimensions.

Since in the limit of large occupancy $D_{col}\varpropto n^2(\bfr,t)$,
we have to discuss non-linear diffusion described by a porous medium equation (PME)\cite{Zeldovich50,CommentPME,Mulansky11,Laptyeva13}:
\be
\partial n/\partial t={\cal D}\nabla (n^{m}\nabla n),
\ee
where $\cal D$ is some constant \cite{CommentPME1}. This equation is simply a continuity equation. The asymptotic solution of the PME is well known \cite{Barenblatt52,Vazquez06}: 
\be
n(r,t)=[(C-kr^{2}/t^{\alpha })/t^{1-\alpha }]^{1/m}>0,
\ee
where $\alpha =2/[dm+2]$, and $C$ and $k$ are constants. We are now interested in $m=2$. Then the obtained expression for $\alpha$ coincides with the exponent $\widetilde{\alpha }=1/(d+1)$ conjectured in Ref.~\onlinecite{Flach09} for spreading of a wave-packet in non-linear disordered lattices in the case when a mechanism of dephasing has been introduced \cite{CommentFlach}. In $d=2$ it yields $\alpha=1/3$.

To get more insight, let us discuss a Bose gas of large occupancy diffusing in a broad stripe. Such a system corresponds to the one-dimensional GPE/NLSE. However, it is natural to expect that the presence of a large number of transverse channels will lead to dephasing. For this case, we get $\alpha=1/2$, which is in full correspondence with the exponent $\widetilde{\alpha}$ observed numerically in Ref.~\onlinecite{Flach09}. Indeed, a sub-diffusive behavior was observed in a cold atom system in the localized regime after adding a repulsive interaction \cite{Lucioni11}.
Note that we can also find the rate of subdiffusion of a wave-packet when the spreading is caused by the force $-\nabla \vartheta$. This yields a current density $j(r,t) \varpropto \frac{n}{\overline{\varepsilon }}D_{col}\nabla \vartheta (r,t)$ which corresponds to $m=3$. In the case of $d=1$, this leads to $\alpha=2/5$, which coincides precisely with the exponent conjectured by Shepelyansky \cite{Shepelyansky93}.

\section{Conclusion}
\label{sec:Conclusion}
In this manuscript, we used the Boltzmann transport equation as a unifying framework for studying the dynamics of the expanding disordered Bose gas. Starting from the Boltzmann equation, we derived a nonlinear kinetic equation for the diffusive regime, including the appropriate collision integral. The collision integral is applicable both for large and for small occupation numbers. A technique of deriving the collision integral starting from the classical GPE was described in our previous paper \cite{Schwiete13}. Here, we recover the result in the limit of large occupations. 

We believe that an analysis in terms of collisional kinetics helps to clarify the physics of sub-diffusive spreading of wave packets in disordered nonlinear lattices. Quite generally, one may guess that the collision-induced diffusion is responsible for the spreading. A properly designed experiment, with a large enough number of Bose atoms and a large number of channels may help to get valuable information about transport properties of disordered interacting systems, both quantum and classical.

\section*{Acknowledgments}
We acknowledge discussions with I.~Gornyi, B.~Shapiro, S.~Flach, T. Kottos, T.~Wellens, K.~Tikhonov, and especially G.~Falkovich. We thank S. Fishman and other participants of the MPAW05 meeting at the Isaak Newton Institute for their interest in  this work. The authors gratefully acknowledge the support by the Alexander von Humboldt Foundation. Both authors thank the members of the Institut f\"ur Theorie der Kondensierten Materie at KIT for their kind hospitality. A.~F. is supported by the National Science Foundation Grant No. NSF-DMR-1006752.

\begin{appendix}
\section{Relation to a recent result on thermalization}
\label{app:therm}
To demonstrate the effectiveness of the kinetic equation approach, we make contact with a work on thermalization of a rarified gas during a stationary diffusion process, Ref.~\onlinecite{Geiger12}. To this end, we specialize on $\mathcal{I}^{LO}$, which contains $\mathcal{F}_{LO}$ of Eq.~\eqref{eq:FLO}  as appropriate for a rarified gas with low occupation numbers. For a steady state process, in the $d=3$ case, the kinetic equation (\ref{eq:diffusion}) can be written as
\be
&&-\left(\nabla -\nabla U\partial_{\eps}-\Gamma_{\eps}\nabla U\right){D}_{\eps}\left(\nabla -\nabla U\partial_{\eps}\right) n_\eps\no\\
&=&2\pi\nu(\eps)\left(\mathcal{I}^{LO}_3\left[n';\mbox{in}\right]-\mathcal{I}^{LO}_3\left[n';\mbox{out}\right]\right).\label{eq:LO}
\ee
Here, we split the collision integral explicitly into "in" and "out" scattering terms, which take the form
\be
\mathcal{I}^{LO}_3\left[n';\mbox{in}\right]&=&\frac{g_3^2}{4\pi}\int d\eps_3 d\eps_4\; \tilde{f}(\eps;\eps_3,\eps_4)\;n_{\eps_3} n_{\eps_4}\no\\
\mathcal{I}^{LO}_3\left[n';\mbox{out}\right]&=&\frac{g_3^2}{4\pi}\int d\eps_2\;\tilde{g}(\eps,\eps_2)\;n_{\eps_2}n_\eps ,
\label{eq:fg}
\ee
where
\be
\tilde{g}(\eps,\eps_2)=\int d\eps_3\; \sqrt{{\min(\eps,\eps_2,\eps_3,\eps+\eps_2-\eps_3)}/{\eps^2\eps_2}},\quad \\
\tilde{f}(\eps;\eps_3,\eps_4)=\sqrt{\min\left(\eps,\eps_3+\eps_4-\eps,\eps_3,\eps_4\right)/\eps\eps_3\eps_4}.\quad
\ee
The above formulas are written under the condition that all energy arguments are larger than zero. Up to the normalization, the functions $\tilde{f}$ and $\tilde{g}$ coincide with the functions $f$ and $g$ obtained in Ref.~\onlinecite{Geiger12} using a diagrammatic approach to describe energy transfer processes.
\end{appendix}


\begin{thebibliography}{51}
\bibitem[{Rem()}]{RemarkCollisions}
\bibinfo{note}{In Ref.~\cite{Schwiete13}, the angular average was not performed
  explicitly for the collision integral.}
\bibitem[{Com({\natexlab{b}})}]{CommentPME}
\bibinfo{note}{We have already discussed the PME with $m=1$ in
  Ref.~\cite{Schwiete10}, in the situation when the nonlinearity was introduced
  by the potential energy.}
\bibitem[{Com({\natexlab{c}})}]{CommentPME1}
\bibinfo{note}{For the disordered Bose gas, the use of the PME is no longer
  justified when occupation numbers become of the order of unity and the
  estimate for the collision rate, Eq.~(27), ceases to be valid.}
\end{thebibliography}
\end{document}